\documentclass[12pt]{iopart}

\usepackage{graphicx,amssymb,array,amsmath}

\newcommand{\bt}{\boldsymbol{t}}
\newcommand{\half}{\mbox{\small $\frac{1}{2}$}}

\begin{document}
\title[Bayesian Model Calibration \& Prediction]{A Bayesian Approach for Parameter Estimation and
  Prediction Using a Computationally Intensive Model}

\author{Dave Higdon$^1$, Jordan D McDonnell$^2$, Nicolas Schunck$^2$, Jason Sarich$^3$,
Stefan M Wild$^3$}

\address{$^1$ Los Alamos National Laboratory, Statistical Sciences, Los Alamos,
NM 87545,
USA}
\address{$^2$ Physics Division, Lawrence Livermore National Laboratory,
Livermore, CA 94551, USA}
\address{$^3$ Mathematics and Computer Science Division, Argonne National
Laboratory, Argonne, IL 60439, USA}
%\address{$^4$
%Department of Physics and Astronomy, University of Tennessee, Knoxville, TN 37996-1200, USA}
%
\ead{dhigdon@lanl.gov}

\begin{abstract}
Bayesian methods have been successful in quantifying
uncertainty in physics-based problems in parameter estimation and prediction.
% No ref numbers allowed in abstract!
%\cite{kaip:some:2004,tarantola2005inverse}.
In these cases, physical measurements $y$ are modeled as the best fit of a physics-based model
$\eta(\theta)$, where $\theta$ denotes the uncertain, best input setting.  Hence the statistical
model is of the form
\[
y = \eta(\theta) + \epsilon,
\]
where $\epsilon$ accounts for measurement, and possibly other, error sources.
When nonlinearity is present in $\eta(\cdot)$, the resulting posterior
distribution for the
unknown parameters in the Bayesian formulation is typically complex and
nonstandard,
requiring computationally demanding computational approaches such as
Markov chain Monte Carlo (MCMC) to produce multivariate draws from the posterior.
Although generally applicable, MCMC requires thousands (or even millions)
of evaluations
of the physics model $\eta(\cdot)$.  This requirement is problematic if the
model takes hours or days
to evaluate.  To overcome this computational bottleneck,
we present an approach adapted from Bayesian model calibration.
% \cite{kenn:ohag:2001}.
This approach combines output from an ensemble of computational model runs with physical
measurements, within a statistical formulation, to carry out inference.
A key component of this approach is a statistical response surface, or emulator, estimated
from the ensemble of model runs.  We demonstrate this approach with a case study in estimating
parameters for a density functional theory model, using experimental
mass/binding energy measurements from a collection of atomic nuclei.
We also demonstrate how this approach produces uncertainties in predictions
for recent mass measurements obtained at
Argonne National Laboratory.
% \cite{van2013first}.
%
\end{abstract}

\pacs{21.10.-k, 21.30.Fe, 21.60.Jz, 21.65.Mn}

\submitto{\JPG}
\maketitle

%%%%%%%%%%%%%%%%%%%%%%%%%%%%%%%%%%%%%%%%%%%%%%%%%%%%%%%%%%%%%%%%%%%%%%%%%%%%%%%
%%%%%%%%%%%%%%%%%%%%%%%%%%%%%%%%%%%%%%%%%%%%%%%%%%%%%%%%%%%%%%%%%%%%%%%%%%%%%%%
%%%%%%%%%%%%%%%%%%%%%%%%%%%%%%%%%%%%%%%%%%%%%%%%%%%%%%%%%%%%%%%%%%%%%%%%%%%%%%%
%%%%%%%%%%%%%%%%%%%%%%%%%%%%%%%%%%%%%%%%%%%%%%%%%%%%%%%%%%%%%%%%%%%%%%%%%%%%%%%

\section{Introduction}
\label{sec:intro}

Bayesian calibration of computer models \cite{kenn:ohag:2001,higd:kenn:cave:2004,baya:berg:paul:2007}
combines output from an ensemble of computational model runs with physical
measurements, within a statistical formulation, in order to carry out
statistical and scientific inferences.
These inferences include quantifying the uncertainty in model parameters as
well as in model-based
predictions.
Unlike more standard Bayesian inverse methodologies
\cite{kaip:some:2004,tarantola2005inverse}, model calibration
approaches must accommodate the high cost of
evaluating the computational model; in many cases, only
a limited number of model evaluations
can be made.
Also, because the model can never exactly match the physical measurements,
even at the best possible parameter input settings, the statistical formulation must
account for this discrepancy.

Since the computational model cannot be quickly evaluated at any input setting
when needed, an ensemble of model runs is carried out prior to
the statistical analysis, producing the raw
material from which to build a response surface, mapping the model
parameter inputs to the model outputs of interest.
A Gaussian process (GP) model is most commonly used to {\em emulate}
the model response as a function of the inputs
\cite{sack:welc:mitc:wynn:1989,oakl:ohag:2004}.  Not only does the GP model
typically produce an accurate emulator \cite{ben2007modeling}, it can
also be embedded within a statistical formulation, allowing parameter
estimation (calibration) and model-based prediction.

In this paper, we apply the Bayesian model calibration approach to the nuclear
density functional theory (DFT) model
described by Schunck et al.\ \cite{Schunck14} in this issue, as well as in
other references
\cite{kortelainen2010nuclear,kortelainen2012nuclear}. To streamline the
presentation, we focus on nuclear
masses for a number of spherical and deformed nuclei as the quantities
of interest for both the model and the experimental measurements.
Our goal is to estimate parameter uncertainties from this data and
to produce predictions, with uncertainty, for a recent set of mass measurements
carried out at the CARIBU facility at Argonne National Laboratory (ANL)
\cite{van2013first}.

In the following sections, we briefly review the experimental
measurements and DFT model before
describing the statistical formulation in greater detail.
We apply the Bayesian model calibration approach to this
data, producing updated uncertainties for the model parameters
and prediction uncertainties for the new ANL mass
measurements,  and we compare these predictions with the experimental
measurements.
%A demonstration on how this emulator-based approach
%can produce uncertainties for the two-neutron drip line is also given.
We end with a discussion of this approach,
pointing out features of this analysis and describing its strengths
and weakness.

%%%%%%%%%%%%%%%%%%%%%%%%%%%%%%%%%%%%%%%%%%%%%%%%%%%%%%%%%%%%%%%%%%%%%%%%%%%%%%%
%%%%%%%%%%%%%%%%%%%%%%%%%%%%%%%%%%%%%%%%%%%%%%%%%%%%%%%%%%%%%%%%%%%%%%%%%%%%%%%
%%%%%%%%%%%%%%%%%%%%%%%%%%%%%%%%%%%%%%%%%%%%%%%%%%%%%%%%%%%%%%%%%%%%%%%%%%%%%%%
%%%%%%%%%%%%%%%%%%%%%%%%%%%%%%%%%%%%%%%%%%%%%%%%%%%%%%%%%%%%%%%%%%%%%%%%%%%%%%%

\subsection{Bayesian Formulations}
\label{sec:bayes}

\begin{figure}[th]
  \centerline{
   \includegraphics[width=6.0in,angle=0] {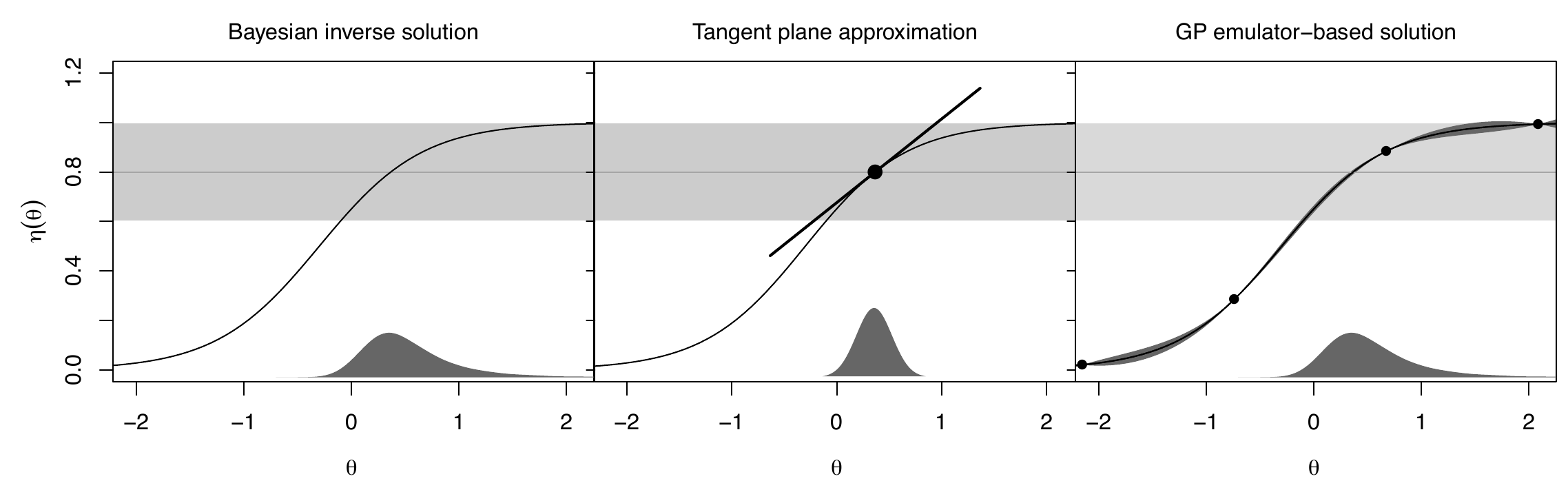}}
   \caption{\label{fig:bayes} Possible Bayesian approximations to a 1-d estimation problem.  The
   horizontal line represents the observation of $y=.8$, whose error has a standard deviation of
   $\sigma=.1$; the $\pm 2\sigma$ region is represented
   by the shaded region about the line.  The sigmoid line shows the model $\eta(\theta)$, and the shaded density on
   the $x$-axis shows the posterior distribution for $\theta$.  Left: the exact
   posterior density for $\theta$ is estimated via MCMC; middle: a linear
tangent approximation to the
   model $\eta(\cdot)$ is used to induce a normal approximation to the
posterior density for $\theta$. Right: a
GP is used to estimate $\eta(\cdot)$ using 4 model runs (black dots), producing a more accurate approximation to the posterior density for $\theta$.}
\end{figure}

Before presenting the Bayesian formulation used for this case study,
we give a brief overview of Bayesian approaches for parameter estimation,
focusing on a simple, 1-d example shown in Figure \ref{fig:bayes}.
Physical data $y$ is combined with a scientifically motivated model $\eta(\cdot)$.  The model requires an input
parameter $\theta$ to make a prediction $\hat{y}=\eta(\theta)$.  The goal is to use the data $y$ to constrain
uncertainty regarding $\theta$.  Uncertainty in $\theta$ then induces
uncertainty in a new prediction $\eta(\theta)$.

The Bayesian paradigm requires specification of the likelihood $L(y|\theta)$, and the prior for the unknown
parameter $\pi(\theta)$.  The inference is based on the posterior distribution, whose (unnormalized) density
is just the product of the likelihood and the prior
\[
 \pi(\theta|y) \propto L(y|\theta) \times \pi(\theta).
\]

In the simple example of Figure \ref{fig:bayes}, the physical observation $y$ is modeled as
\[
 y = \eta(\theta) + \epsilon,\mbox{ where } \epsilon \sim N(0,0.1^2),
\]
where $\sim$ means ``is distributed as.''
Thus $L(y|\theta) \propto \exp\{-\half \cdot 0.1^{-2} (y - \eta(\theta))^2 \}$.
For the prior
we take $\theta \sim N(0,1)$ so that $\pi(\theta) \propto \exp\{-\half \theta^2\}$.
Thus the resulting posterior density for $\theta$ is given by
\begin{equation*}
\label{eq:simplePost}
\pi(\theta|y) \propto \exp\left\{-\half \left[ 0.1^{-2} (y - \eta(\theta))^2 +
\theta^2\right] \right\}.
\end{equation*}

Bayesian inference requires understanding this resulting posterior distribution
and using it to make
predictions for new observations $y$.  While the 1-d example shown here is not too daunting, typical examples
have a much higher-dimensional parameter space, including statistical nuisance parameters
such as variances.  In such cases, Markov chain Monte Carlo (MCMC)
\cite{gamerman2006markov} can be used to generate a realization from an
ergodic Markov chain, producing a (dependent)
sequence of samples $\{\theta^{(1)},\ldots,\theta^{(T)}\}$ from the posterior distribution $\pi(\theta|y)$.
This Markov chain is most commonly produced using some form of rejection step, where the
transition from $\theta^{(k)}$ to $\theta^{(k+1)}$ is taken by proposing a new value
$\theta^*$, and setting $\theta^{(k+1)}$ to either $\theta^*$ or $\theta^{(k)}$
with a probability that depends on the ratio
$\pi(\theta^*|y)/\pi(\theta^{(k)}|y)$.
Hence, producing a useful sample from $\pi(\theta|y)$ with
MCMC requires many thousands to millions of evaluations of $\eta(\theta)$.
In our case study, $\eta(\theta)$ represents a computationally
demanding DFT model, making direct use of MCMC infeasible.  A common alternative is to
replace the computational model $\eta(\theta)$ by a linear tangent model approximation.
While easy to work with, this linear assumption often results in overly narrow estimates of the
parameter uncertainty
(see the middle frame of Figure \ref{fig:bayes}).  In this paper we develop a
Gaussian process (GP)-based approximation
for $\eta(\cdot)$, allowing posterior sampling with MCMC, without additional
evaluations of $\eta(\theta)$.

%%%%%%%%%%%%%%%%%%%%%%%%%%%%%%%%%%%%%%%%%%%%%%%%%%%%%%%%%%%%%%%%%%%%%%%%%%%%%%%
%%%%%%%%%%%%%%%%%%%%%%%%%%%%%%%%%%%%%%%%%%%%%%%%%%%%%%%%%%%%%%%%%%%%%%%%%%%%%%%

\subsection{DFT Model and Experimental Measurements }
\label{sec:dft}

The DFT model considered in this case study is based on the UNEDF1
parametrization of the Skyrme functional \cite{kortelainen2012nuclear}. It
requires a $p=12$-dimensional
parameter vector
$t = (t_1,\ldots,t_p)$ to produce fitted masses for any given nucleus.
The parameters and their prior ranges are given in Table \ref{tab:params}.  For any
input setting $t$ and any specified nucleus $(N,Z)$, the DFT code produces
a fitted mass.
Note that we use $t$ to denote a generic setting of the inputs, reserving $\theta$ to denote
the unknown best input setting, which is to be estimated from the experimental
mass measurements.

\begin{figure}[ht]
  \centerline{
   \includegraphics[width=6.0in,angle=0] {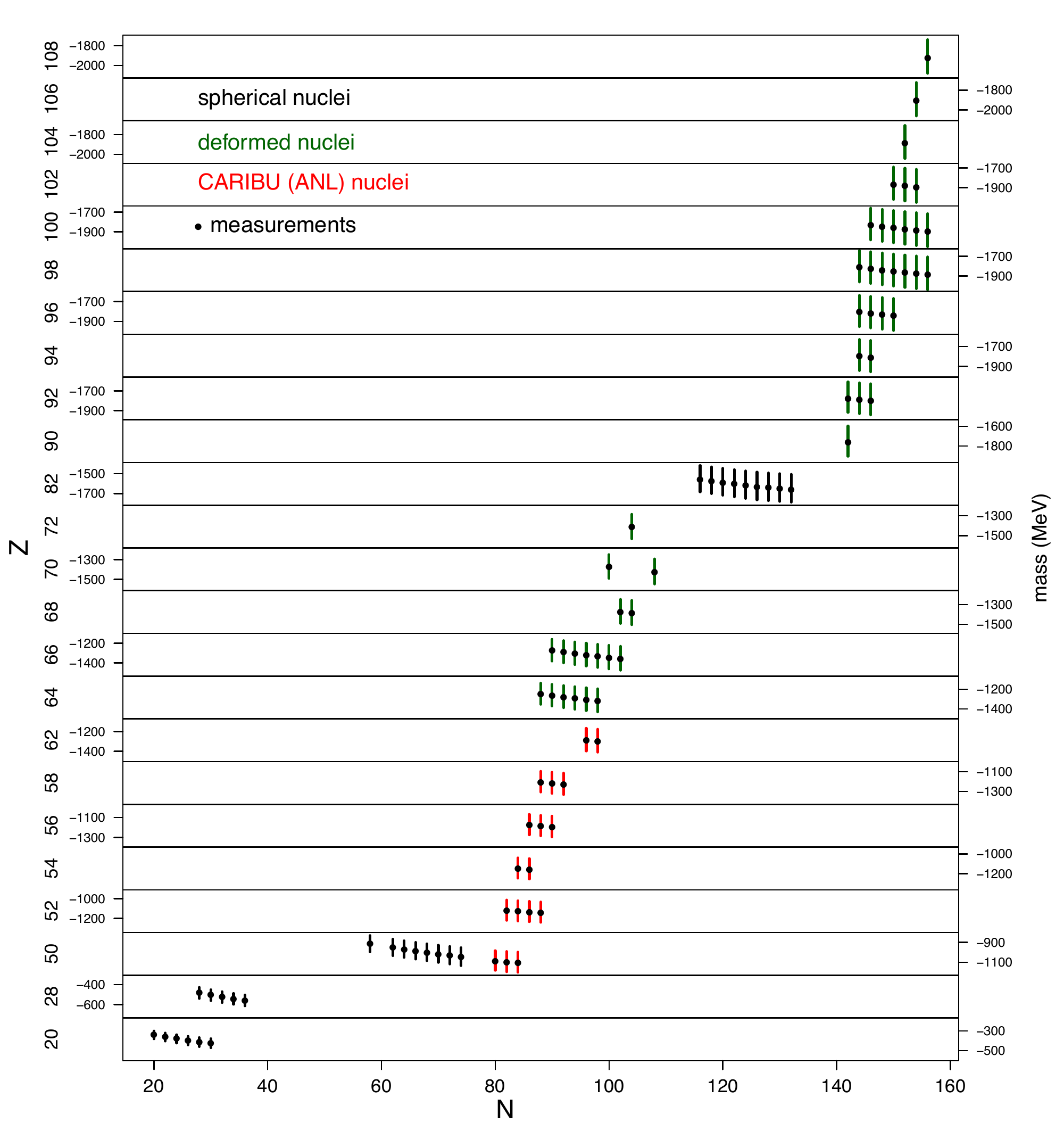}}
   \caption{\label{fig:mass} Range of DFT computed masses (vertical lines),
   along with the experimental measurements (dots) for the 28 spherical (black)
   and 47 deformed (green) nuclei of the original UNEDF1 parametrization, as well
   as the 17 newly measured neutron-rich nuclei (red). The ranges of DFT
   computed masses are derived from the ensemble of 183 parameter settings
   shown in Figure \ref{fig:design}.}
\end{figure}

\begin{table}[ht]
\caption{\label{tab:params} DFT model parameters for the UNEDF1 parametrization
and their prior ranges\\[-.2cm]}
\centerline{
\begin{tabular}{cccc}
\hline \hline
Parameter & Label                & $C_{\rm lower}$ & $C_{\rm upper}$\\
\hline
$\rho_c$                   &    $\theta_1$ &  0.155 & 0.165 \\
$E^{NM}/A$               &  $\theta_2$ & $-16.0$ &  $-15.5$ \\
$K^{NM}$                   & $\theta_3$ & 200 & 240 \\
$a^{NM}_{\mathrm{sym}}$ &  $\theta_4$ &  27.0 & 31.0\\
$L^{NM}_{\mathrm{sym}}$    &   $\theta_5$ & 15.0 & 65.0\\
$1/M_s^*$                  &    $\theta_6$ & 0.75 & 1.25\\
$C^{\rho\Delta \rho}_0$    &  $\theta_7$ & $-60$ &  $-30$\\
$C^{\rho\Delta\rho}_1$     & $\theta_8$ & $-240$ &  $-50$ \\
$V^n_0$                    & $\theta_9$ & $-220$ &  $-150$\\
$V^p_0$                    & $\theta_{10}$ & $-230$ & $-180$\\
$C^{\rho\nabla J}_0$       & $\theta_{11}$ & $-90$ & $-60$\\
$C^{\rho\nabla J}_1$       &  $\theta_{12}$ & $-90$ &  $ 20$   \\[.1cm]
\hline \hline
\end{tabular}
}
\end{table}
The example illustrated in Figure \ref{fig:mass} considers the masses of the 28
spherical nuclei and 47 deformed nuclei included in the original UNEDF1
parametrization, supplemented by the masses of 17 neutron-rich nuclei recently
measured at the CARIBU facility at Argonne National Laboratory \cite{van2013first}.
For a given
input setting $t$, the DFT code is
run to compute the masses for all $n = n_1+n_2+n_3 = 28+47+17 = 92$ nuclei, producing an $n$-vector
of outputs $\eta(t)$.  For the Bayesian analysis, we generate an initial set of $m=183$
DFT model runs for each of these $n$ nuclei.  Note that the original design was for 200 parameter
settings, but 17 of these runs were discarded because of convergence issues.
The output range of the resulting $n \times m = 16,836$ DFT runs
are shown in Figure \ref{fig:mass} for
each of the $92$ nuclei.
The input settings used in this analysis are shown
in Figure \ref{fig:design}.

In addition to the model runs, each nucleus has an experimentally determined
mass denoted by the
black dot in Figure \ref{fig:mass}.  As a demonstration of this Bayesian model calibration methodology,
we use the data from the 28 spherical and 47 deformed nuclei of UNEDF1 to constrain
model parameter uncertainties
and to estimate model error.  With these results, we compare the resulting
predictions, and their uncertainties,
with the actual measurements of the 17 neutron-rich nuclei obtained at ANL \cite{van2013first}.

%%%%%%%%%%%%%%%%%%%%%%%%%%%%%%%%%%%%%%%%%%%%%%%%%%%%%%%%%%%%%%%%%%%%%%%%%%%%%%%
%%%%%%%%%%%%%%%%%%%%%%%%%%%%%%%%%%%%%%%%%%%%%%%%%%%%%%%%%%%%%%%%%%%%%%%%%%%%%%%

\section{Bayesian Formulation}

The full Bayesian model formulation combines experimental measurements and an ensemble of DFT
model runs, all within an encompassing statistical model.
% This encompassing model formulation is rather involved.
Hence we describe the
main components of this model in the next three subsections.
The first component, described in Section \ref{sec:emu}, is the GP model used to
probabilistically describe the DFT output given parameter settings $t$.  The second component,
described in Section \ref{sec:modcal}, is
a full Bayesian model calibration formulation for a single output type (spherical masses),
using the simulation output, along with the
experimental measurements to reduce parameter uncertainties.
The third component, described in Section \ref{sec:modcal2}, combines separate
formulations into a single Bayesian
model, so that information from multiple data types can be used to estimate
parameter uncertainties -- both for the DFT model and for statistical
parameters that
control the GP covariance and the error variance.

Once the various parameter uncertainties are estimated within this overarching statistical
model, predictions and predictive distributions for outputs of interest can be determined.
We will compare the predictive distribution with each of the experimental
measurements used
for estimation (i.e., the masses for the spherical and deformed nuclei).  We
will
also compare
predictive distributions for the new ANL mass measurements for neutron-rich nuclei.
Since the ANL mass measurements were not used to estimate any of the model parameters,
we can assess the quality of the predictions produced by this
statistical formulation.

\subsection{Emulation of DFT Model Output}
\label{sec:emu}
The DFT model requires 5--10 minutes to compute the mass for a given nucleus.
Given that the
statistical analysis presented here involves 75 nuclei, requiring 75 DFT solves for every evaluation of
$\eta(\theta)$, direct use of MCMC is not practical.  In this case study,
therefore, we treat $\eta(\cdot)$ as an
unknown function, to be estimated from an initial set of $m=183$ DFT solves for each nucleus.  In this
section we describe a Bayesian approach for estimating $\eta(\cdot)$ from the $m$ training runs with
a Gaussian process (GP) specification. A Bayesian approach is taken that can be integrated into the
full model formulation.

The use of a GP model to {\em emulate} a computational model $\eta(\cdot)$ at new parameter inputs
dates back $25+$ years \cite{sack:welc:mitc:wynn:1989,welc:buck:sack:wynn:mitc:morr:1992}.
The approach has proven effective in various applications where the model
output
changes smoothly as a function of the
inputs $\theta$ \cite{o2006bayesian,heit:higd:habi:2006,sanso2008ics}.

\subsubsection*{The Ensemble of DFT Runs.}

\begin{figure}[ht]
  \centerline{
   \includegraphics[width=5.5in,angle=0] {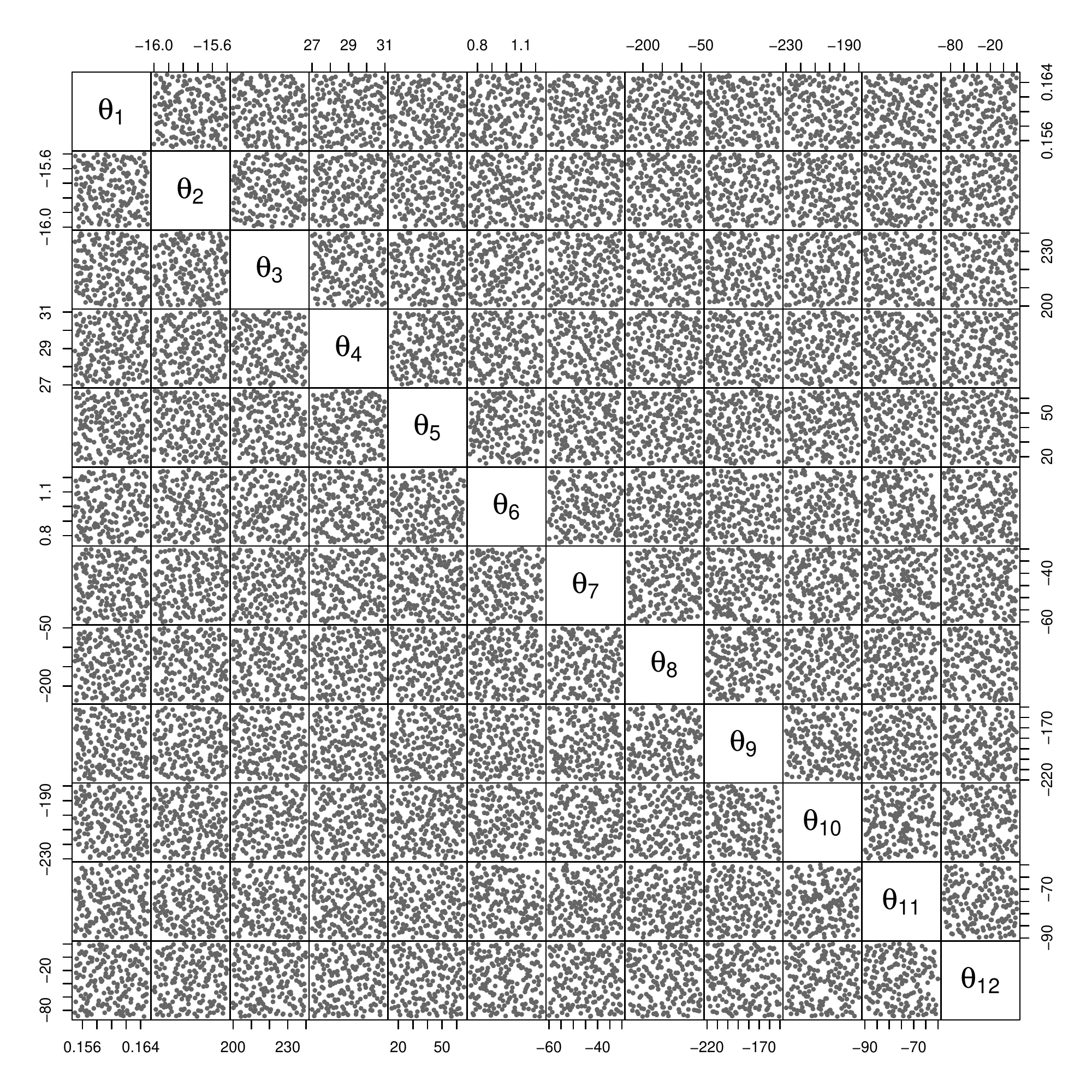}}
   \caption{\label{fig:design} Space-filling Latin hypercube sample (LHS) used
to specify the initial
   set of DFT runs with which to construct the GP emulator.
   The 2-d projections of this design are shown for each pair of parameters.}
\end{figure}

We start with simulator runs at $m$ different input settings
\[
  \eta(t^*_j),\; j=1,\ldots,m.
\]
We use a space-filling Latin hypercube \cite{Tang:1993} sample for this initial design of input settings $(t_1^*,\ldots,t_m^*)$.
We use $\bt^*$ to denote the $m \times p$ matrix describing this design, or ensemble of input settings.
Two-dimensional projections of this parameter design are shown in Figure \ref{fig:design}.
How best to construct a design for an emulator is still a research topic in the statistical literature; a starting point to
this literature can be found in Santer et al.'s textbook \cite{Sant:Will:Notz:2003} and the references therein.

\subsection{Basis Representation of the DFT Model Output}

We focus here on a single output type: the $n=28$ masses
for the spherical nuclei.  There are a number of approaches available
for extending the univariate GP model to handle many outputs at once
\cite{rougier2008efficient,bayarri2009using}; we use the
basis approach of \cite{higdon2008cmc} which performs well in this application.

To simplify the model specification, we standardize the $p=12$-dimensional parameter
space to $C = [0,1]^p$ using the ranges given in Table \ref{tab:params}.
%as well as the model output so that the means and variance of
%the ensemble of masses are 0 and 1 respectively.
%
For a given input $t$ in the standardized input space $[0,1]^{p}$, the DFT model produces an $n$-vector $\eta(t)$, giving
the mass for each of the $n$ nuclei.  The emulator models the DFT output using a $q$-dimensional basis representation:
\begin{equation}
\label{eq:etabasis}
 \eta(t) = \sum_{i=1}^{q} \phi_i w_i(t) + e,\;
 t \in [0,1]^{p},
\end{equation}
where  $\{\phi_1,\ldots,\phi_{q}\}$ are orthogonal, $n$-dimensional basis
vectors, the $w_i(t)$ are weights whose
value depends on $t$, and $e$ is an $n$-dimensional error term, accounting for
the residual in the basis
representation, as well as numerical noise
 -- associated with finite-precision (e.g., roundoff error) and finite-process
(e.g., residual tolerances, adaptive discretizations) computations --
from the DFT code.
This formulation builds an emulator that maps $[0,1]^{p}$ to $R^{n}$ by building
$q$ independent, univariate models for each $w_i(t)$. Separate Gaussian
processes models \cite{sack:welc:mitc:wynn:1989,higdon2008cmc} are used to model
each of the weight functions; this is described in the following two subsections.
%The details of this model specification are given below.

For each of the $m=183$ settings $(t^*_1,\ldots,t^*_m)$ of parameter design, an $n$-dimensional vector of masses is produced,
giving $\eta_1,\ldots,\eta_m$.
These output vectors are represented by principal components
\cite{rams:silv:1997} or, equivalently, by empirical orthogonal functions (EOFs)
\cite{eof:1999}.  Following standard practice, the output vectors are centered
by subtracting the
mean ($\frac{1}{m} \sum_{j=1}^m \eta_j$) from each output vector.  Alternative standardizations may be preferred,
depending on the application.  This same standardization is also applied to the experimental data.

\begin{figure}[ht]
  \centerline{
   \includegraphics[width=6in] {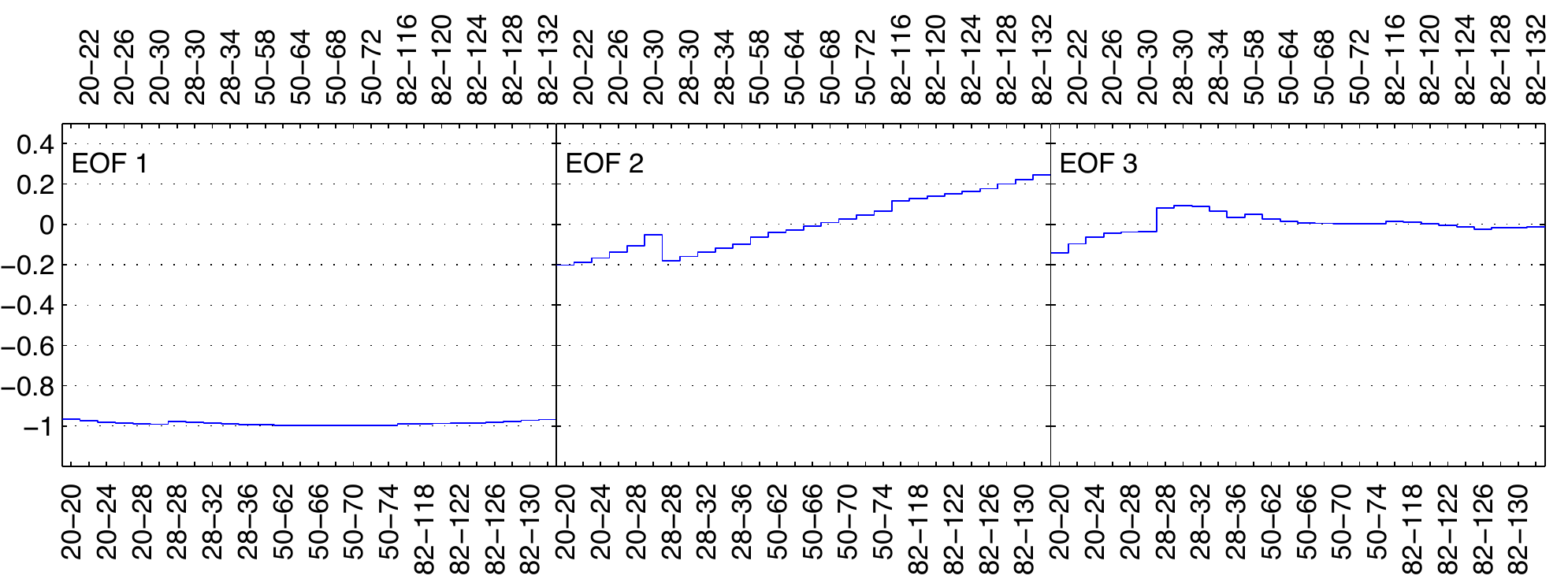}}
   \caption{\label{fig:eofs} The first three basis functions $\phi_i$, $i=1,\ldots,3$, over each of the 28 spherical nuclei.
   Here each of the basis functions is derived by using empirical orthogonal
functions from the 183 simulations.
   Each nucleus is labeled $Z$-$N$.}
\end{figure}

We obtain the $n \times m$ matrix $\Xi$ by column-binding the (standardized) output vectors from the simulations.
%\begin{equation}
%\Xi = [\eta_1;\cdots;\eta_m].
%\end{equation}
Applying a singular value decomposition (SVD) to the simulation output matrix
$\Xi$ gives
\begin{equation}
 \Xi = [\eta_1;\cdots;\eta_m] = UDV',
\end{equation}
where $U$ is a $n \times m$ orthogonal matrix, $D$ is a diagonal $m \times m$ matrix holding the singular values, and $V$ is a $m \times m$ orthonormal matrix. To construct a $q$-dimensional representation of the simulation output, we define the EOF basis matrix $\Phi_\eta$ to be the first $q$ columns of $[UD\sqrt{m}]$.  We take $q = 9$; this is sufficient to explain over 99.9\% of the variation in the simulation ensemble. For illustration, the first three basis functions $\phi_1, \ldots, \phi_3$ are
shown in Figure  \ref{fig:eofs}.
Note that the $\phi_i$ are $n=28$-dimensional vectors, with one element for
each spherical nucleus.

\subsubsection{Specifying the Gaussian Process Model Emulator.}

Each of the basis weights in Equation (\ref{eq:etabasis}), $w_i(t)$, $i=1,\dots,q$, is
a function mapping the $p$-dimensional input $t$ to a scalar.  These functions are
modeled (a priori) as
independent, mean zero GPs,
\begin{equation}
%\label{eq:gpw}
  w_i(\cdot) \sim
\mbox{GP}(\mu(\cdot),C(\cdot,\cdot))  \mbox{ with }\mu(t)=0,\,C(t,t') = \lambda^{-1}_{wi} R(t,t';\rho_{i}),
\end{equation}
where $\lambda_{wi}$ is the marginal precision (precision = 1/variance) of the process and
$R(t,t';\rho_{i})$ is a correlation function, whose entries depend on the pair
of input settings $t$ and $t'$, as well as the vector $\rho_i =
(\rho_{i1},\ldots,\rho_{ip})'$:
\begin{equation}
\label{eq:wcor}
R(t,t';\rho_i) = \prod_{k=1}^{p} \rho_{ik}^{4(t_{k} - t'_{k})^2}.
\end{equation}
This is the Gaussian covariance function, giving smooth realizations, and
commonly used to model computer simulation output
\cite{kenn:ohag:2001,sack:welc:mitc:wynn:1989}.  This fits well with our
expectation that the DFT masses change smoothly as the input values are
changed.  An advantage of the product form is that only a single additional
parameter is required per additional input dimension, while the fitted GP
response still allows for general interactions between inputs.
The parameter $\rho_{ik}$ controls the spatial range for the $k$th input
dimension of the process $w_i(\cdot)$. Under this parameterization, $\rho_{ik}$
gives the correlation between $w_i(t)$ and $w_i(t')$ when the input conditions
$t$ and $t'$ are identical, except for a difference of 0.5 in the $k$th
component.  Note that this interpretation uses the standardization of the
input space to $[0,1]^{p}$.

Restricting our attention to the $m$ input design settings, we define the
$m$-vector $w_i$ to be $w_i = (w_i(t^*_{1}),\ldots, w_i(t^*_{m}))'$ for
$i=1,\ldots,q$.
In addition we define $R(\bt^*;\rho_{i})$ to be the $m \times m$ correlation
matrix resulting from applying Equation (\ref{eq:wcor}) to each pair of input
settings in the design $\bt^*$.  The $p$-vector $\rho_{i}$ gives the correlation
distances for each of the input dimensions.  At the $m$ simulation input
settings, the $m q$-vector $w=(w_1',\ldots,w_{q}')'$ then has prior distribution
\begin{equation}
\label{eq:wprior1}
%  w=
  \begin{pmatrix} w_1 \cr \vdots \cr w_{q} \end{pmatrix}
  \sim
  N\left( \begin{pmatrix} 0 \cr \vdots \cr 0 \end{pmatrix},
  \begin{pmatrix}
     \lambda^{-1}_{w1}R(\bt^*;\rho_{1})
                                      & 0 & 0 \cr
     0 & \ddots & 0 \cr
     0 & 0 & \lambda^{-1}_{wq}R(\bt^*;\rho_{q})
  \end{pmatrix}
  \right),
\end{equation}
which is controlled by $q$ precision parameters held in $\lambda_{w}$ and $q
\cdot p$ spatial correlation parameters held in $\rho$. The prior above can be
written more compactly as $w \sim N(0,\Sigma_w)$, where $\Sigma_w$, controlled
by parameter vectors $\lambda_w$ and $\rho$, is given by the block diagonal
covariance matrix in Equation (\ref{eq:wprior1}).

\subsubsection{Bayesian Representation of the GP Emulator.}

We specify independent gamma priors $Ga(a_w,b_w)$ for each $\lambda_{wi}$ and
independent beta priors for the $\rho_{ik}$,
giving the prior densities
\begin{eqnarray} \label{eq:priors}
\pi(\lambda_{wi}) & \propto & \lambda_{wi}^{a_w-1} e^{-b_w \lambda_{wi}},
   \;\; i=1,\ldots,q, \\
\pi(\rho_{ik}) & \propto & \rho_{ik}^{a_{\rho}-1} (1-\rho_{ik})^{b_{\rho}-1},
   \;\; i=1,\ldots,q, \, k=1,\ldots, p.  \nonumber
\end{eqnarray}
We expect the marginal variance for each $w_i(\cdot)$ process to be close to
one, because of the scaling of the basis functions.  For this reason we specify
that $a_w
= b_w= 5$, encouraging each $\lambda_{wi}$ to be close to 1.  In addition, this
informative prior helps stabilize the resulting posterior distribution for the
correlation parameters that can trade off with the marginal precision
parameter.  Because we expect only a subset of the inputs to influence the
simulator response, our prior for the correlation parameters reflects this
expectation of {\em effect sparsity} for each $w_i(\cdot)$.
Under the parameterization in Equation
(\ref{eq:wcor}), input $k$ is inactive for PC $i$ if $\rho_{ik}=1$.  Choosing
$a_{{\rho}}=1$ and $0 < b_{{\rho}} < 1$ yields a density with substantial
prior mass near 1.  We take $b_{\rho} = 0.1$, which makes Pr$(\rho_{ik} <
0.98) \approx \frac{1}{3}$ a priori.  In general, the selection of these
hyperparameters should depend on how many of the $p$ inputs are expected to be
active.
%Alternatively, prior for $\rho_{ik}$ could some point mass at one to encourage effect sparsity \cite{link:bing:heng:2006}.

We can now define the likelihood, or sampling model, for the simulation output $\eta$.
Here  $\eta = \mbox{vec}(\Xi)$, where vec($\Xi$) produces a vector by stacking the columns of matrix $\Xi$.   Taking the error vector in Equation (\ref{eq:etabasis}) to be independent Gaussian with common precision $\lambda_\eta$, we get the sampling model,
or likelihood, for $\eta$:
\begin{equation}\label{eq:etagivenw}
\eta|w,\lambda_\eta \sim N \left( \Phi w, \lambda_\eta^{-1} I \right),
\end{equation}
where  $\Phi = [I_m \otimes \phi_1; \cdots ; I_m \otimes \phi_{q}]$ and the
$\phi_i$ are the $q$ basis vectors previously computed by SVD.  A
$Ga(a_\eta,b_\eta)$ is specified for the error precision $\lambda_\eta$.

Multiplying the probability density functions implied by Equations
(\ref{eq:wprior1}), (\ref{eq:priors}), and (\ref{eq:etagivenw}) and the gamma
prior for $\lambda_\eta$ yields the (unnormalized) posterior density.
After integrating out $w$, the posterior distribution for the unknown parameters becomes
\begin{eqnarray}
\label{eq:postw}
\lefteqn{ \pi(\lambda_\eta,\lambda_w,\rho| \eta)  \propto } \nonumber \\
  & & \left| (\lambda_\eta \Phi'\Phi)^{-1} + \Sigma_w \right|^{-\frac{1}{2}}
  \exp\{ -\half \hat{w}' ([\lambda_\eta \Phi'\Phi]^{-1} + \Sigma_w)^{-1} \hat{w} \}
  \times \\
\nonumber
  & &
  \lambda_\eta^{a^*_\eta-1} e^{-b^*_\eta \lambda_\eta} \times
  \prod_{i=1}^{q} \lambda_{wi}^{a_w-1} e^{-b_w \lambda_{wi}} \times
  \prod_{i=1}^{q}
       \prod_{j=1}^{p} (1-\rho_{ij})^{b_\rho-1},
\end{eqnarray}
where
\begin{eqnarray}
a^*_\eta &=& a_\eta+\frac{m(n-q)}{2},\nonumber \\
b^*_\eta &=& b_\eta + \half \eta' (I-\Phi(\Phi'\Phi)^{-1}\Phi') \eta,\mbox{ and}\\
\hat{w} &=& (\Phi'\Phi)^{-1}\Phi' \eta. \nonumber
\end{eqnarray}

In some applications $\hat{w}$ -- the simulation output dotted with the basis vectors -- may not exactly
conform to a smooth response over the input space.  This situation is often due
to numerical jitter in the computational model.
In such cases, an additional error may be required.
We typically add the $mq \times mq$ covariance matrix $\mbox{diag}(\lambda_{o1}I_m,\ldots,\lambda_{oq}I_m$ to
the covariance term in (\ref{eq:postw}) to allow for some mismatch between
$\hat{w}_i$ and $w(t_i)$.  In this case,
independent $Ga(1,.0001)$ priors are used for the $\lambda_{oi}$'s.

\subsubsection{Exploring the Posterior Distribution for the Emulator.}

\begin{figure}[ht]
\centerline{
 \includegraphics[width=4.7in,angle=0,clip=]{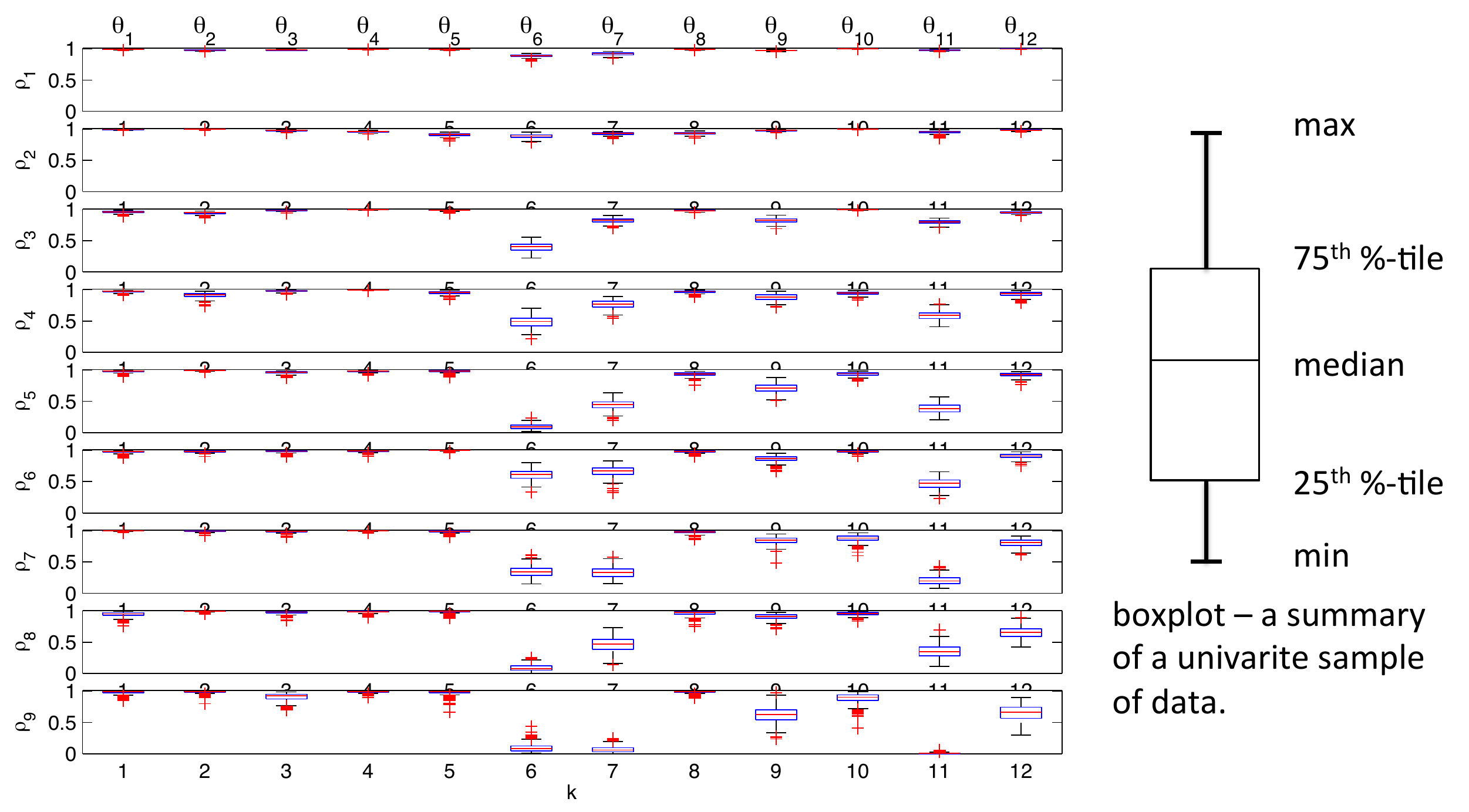}
}
\caption{ Boxplots of posterior samples for each $\rho_{ik}$, which control the GP response surface that predicts
mass for the spherical nuclei as a function of the 12 DFT parameters.
%A boxplot summarizes a univariate sample
%of data. The box range depicts the middle 50\%
%of the data; the whiskers extending from the box show the range of the remaining data.
\label{fig:rhobox}}
\end{figure}

\begin{figure}[th]
\centerline{
 \includegraphics[width=5.2in,angle=0,clip=]{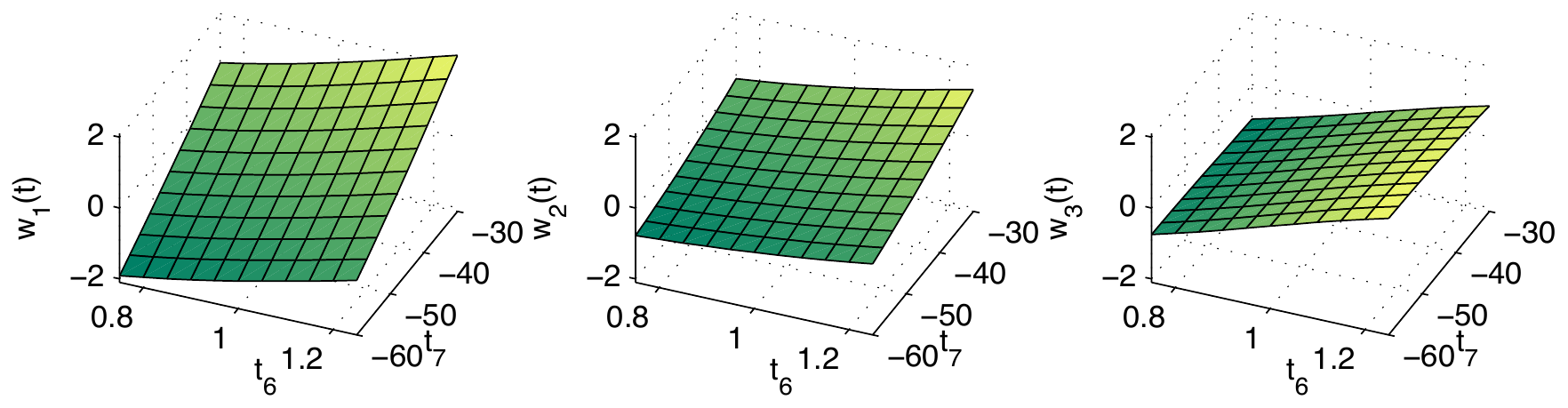}
}
\caption{ Posterior mean surfaces for $w_i(\cdot),\,i=1,2,3$, the weights corresponding to the first three EOF's.
Here the remaining 10 parameters were held at their midpoints as $t_6$ and $t_7$ vary over their design range. \label{fig:wpred}}
\end{figure}

The posterior distribution is a necessary ingredient for the complete
formulation
that incorporates the experimental data.  However, it is often also worth
exploring this intermediate posterior distribution for the DFT model response.
For this purpose we use MCMC and standard Metropolis updates
\cite{BGHM:95,gamerman2006markov}
and we view a number of posterior quantities to illuminate features of the DFT
model.
The posterior of the emulator response can be used to investigate sensitivity measures of computational model
\cite{oakl:ohag:2004} or to estimate a Sobol decomposition of the model response
\cite{sack:welc:mitc:wynn:1989}.
Figure  \ref{fig:rhobox} shows boxplots of the posterior distributions for the components of $\rho$.  From this figure it is
apparent that the PC's are influenced by a number of the components in $t$.
Figure \ref{fig:wpred} shows the resulting posterior mean surfaces for
$w_1(\cdot)$, $w_2(\cdot)$,  and $w_3(\cdot)$ as a function of $t_6$ and $t_7$.

\subsubsection{Generating Emulator-Based Predictions.}

Given the posterior realizations from Equation (\ref{eq:postw}), one can generate realizations from the process $\eta(\cdot)$ at any input setting $t^\star$.
Since
\begin{equation}
\eta(t^\star) =
  \sum_{i=1}^{q} \phi_i w_i(t^\star),
\end{equation}
realizations from the $w_i(t^\star)$ processes need to be drawn given the MCMC output.  For a given draw $(\lambda_\eta,\lambda_w,\rho)$ a draw of
$w^\star = (w_1(t^\star),\ldots,
            w_{q}(t^\star))'$
can be produced by using the fact
\begin{equation}
 \begin{pmatrix} \hat{w} \cr w^\star \end{pmatrix}
 \sim
 N \left( \begin{pmatrix} 0 \cr 0 \end{pmatrix},
   \left[ \begin{pmatrix} (\lambda_\eta \Phi'\Phi)^{-1} & 0 \cr 0 & 0 \end{pmatrix}
   + \Sigma_{w,w^\star}(\lambda_w,\rho) \right] \right),
\end{equation}
where $\Sigma_{w,w^\star}$ is obtained by applying the covariance rule from Equation (\ref{eq:wcor}) to the augmented input settings that include the original design $\bt$ and the new input setting $t^\star$.  Recall that $\hat{w} = (\Phi'\Phi)^{-1}\Phi' \eta$.  Application of the conditional normal rules then gives
\begin{equation}
\label{eq:wdist}
  w^\star|\hat{w} \sim N(V_{21}V_{11}^{-1}\hat{w},
                 V_{22}-V_{21}V_{11}^{-1}V_{12}),
\end{equation}
where
\begin{equation}
  V =
  \begin{pmatrix} V_{11} & V_{12} \cr V_{21} & V_{22} \end{pmatrix} =
  \left[ \begin{pmatrix} (\lambda_\eta \Phi'\Phi)^{-1} & 0 \cr 0 & 0 \end{pmatrix}
   + \Sigma_{w,w^\star}(\lambda_w,\rho) \right]
\end{equation}
is a function of the parameters produced by the MCMC output.  Hence, for each
posterior realization of $(\lambda_\eta, \lambda_w, \rho)$, a realization of
$w^\star$ can be produced.  This approach easily generalizes to give
predictions over many input settings at once.

\begin{figure}[ht]
  \centerline{%\fbox{
   \includegraphics[width=6in] {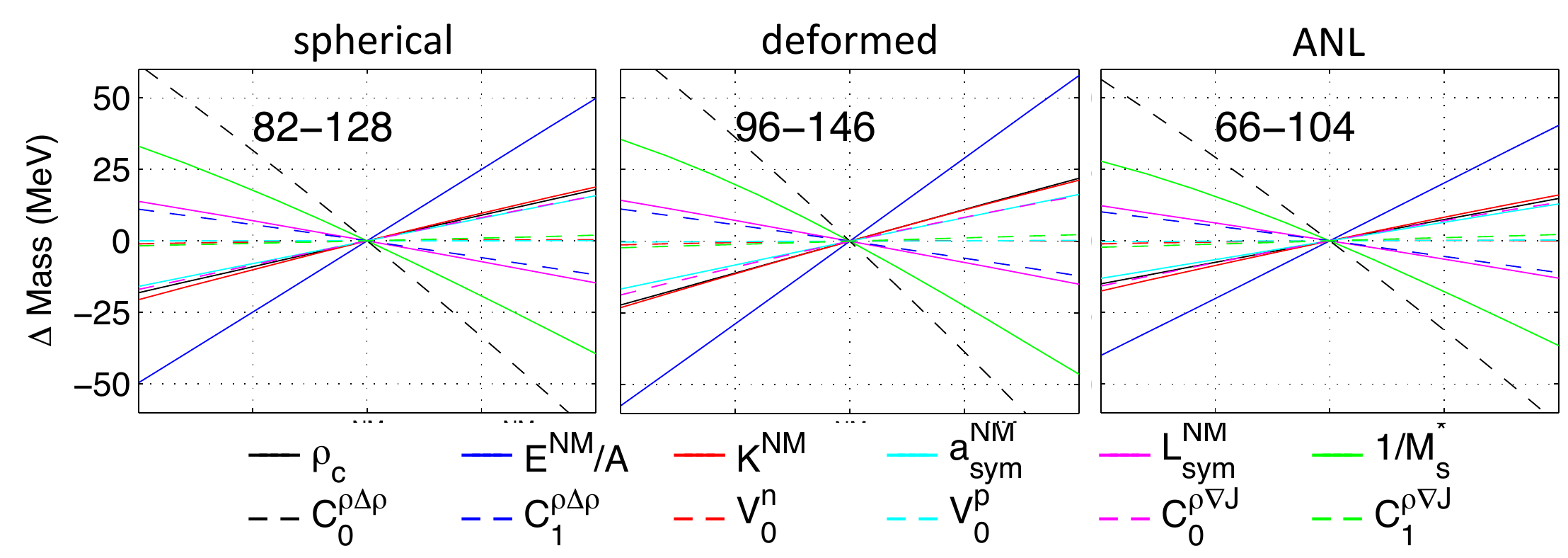}}%}
   \caption{\label{fig:sens1} Sensitivity of the masses computed via DFT as each
of the 12 parameters is varied from low to high (for ranges see Table
\ref{tab:params}).  The plots show how computed masses change for three
nuclei (Z-N) -- one
   spherical, one deformed, and one from the new ANL measurements -- as the parameters are varied, one at at time.}
\end{figure}

Figure \ref{fig:sens1} shows posterior means for the simulator response $\eta(\cdot)$ where each of the inputs is varied over its prior (standardized) range of $[0,1]$ while the other 11 inputs are held at their midpoints.

The posterior mean response conveys an idea of how the different parameters affect the computed masses for these
three nuclei.   The sensitivities for other spherical, deformed, and
neutron-rich ANL masses are similar.
Other marginal functionals of the simulation response can also be calculated,
such as
sensitivity indices or estimates of the Sobol decomposition
\cite{sack:welc:mitc:wynn:1989,oakl:ohag:2004}.

Note that a simpler emulator could have been constructed by
estimating $(\lambda_\eta, \lambda_w, \rho)$ via
the posterior mean or via maximum likelihood.  Conditional on these parameters,
the model output
could be emulated by using the mean in Equation (\ref{eq:wdist}).
Also, we opted to build the emulator over the 12-d parameter space $C$, requiring
the DFT model to predict at any $(Z,N)$, rather than emulate over this
extended $12+2$-d space.  This was decided because we expected less continuity
across the $(Z,N)$ space.  However, assessing the relative strengths and weaknesses
of alternative emulators is not well studied for this application.  Our EOF-based
GP emulator is sufficiently accurate for this application; testing
against holdout DFT runs for randomly chosen $t \in C$ yields a root mean square error
of 0.14 and 0.17 MeV for spherical and deformed masses respectively.

%%%%%%%%%%%%%%%%%%%%%%%%%%%%%%%%%%%%%%%%%%%%%%%%%%%%%%
%%%%%%%%%%%%%%%%%%%%%%%%%%%%%%%%%%%%%%%%%%%%%%%%%%%%%%

\subsection{Full Bayesian GP-based Formulation -- Single Output Type}
\label{sec:modcal}

We now describe the model formulation that incorporates experimental measurements
to constrain uncertainty regarding the DFT parameter vector $\theta$.  We focus
on the
spherical mass data and emulator; in the following subsection, we describe how
the multiple data types
can be combined in a common formulation.  Hence we use $n$ to denote the number
of experimental measurements, $m$ to denote the number of model runs for each
nucleus, and
$(\lambda_y,\lambda_\eta,\lambda_w,\rho)$ to denote the model parameters specific to this data
type, leaving the distinction between data types to be made in the following subsection.

We have mass measurements for $n=28$ spherical nuclei held in the $n$-vector
$y$.  Although these
measurements are accurate, we expect errors $\epsilon$ between the measurements
and the DFT code, even
at the (unknown) best setting $\theta$, giving
\begin{equation}
\nonumber
 y=\eta(\theta) + \epsilon,
\end{equation}
where the errors are modeled as $N(0,\Sigma_y)$.   We define $\Sigma_y =
\lambda_y^{-1} I_n$ and
specify a diffuse $Ga(a_y=1,b_y=.005)$ prior for $\lambda_y$, allowing the experimental measurements
to inform about these precisions.  Using the basis representation for
$\eta(\cdot)$ in (\ref{eq:etabasis}), we obtain
a normal-gamma form for the data model
\begin{equation}
\label{eq:datalike}
 y| w(\theta), \lambda_y \sim
     N( \Phi w(\theta),\Sigma_y),\,\,\,
 \lambda_{y} \sim Ga(a_y,b_y).
\end{equation}

We can now write out the entire posterior distribution for all the parameters,
including $\theta$.  First, let
\begin{eqnarray}
  \hat{w}_y &=& (\Phi' \Sigma^{-1}_y \Phi)^{-1} \Phi' \Sigma^{-1}_y y, \nonumber \\
    a^*_{y} &=& a_y + \half (n -  q), \nonumber \\
    W_{y} &=& \lambda_{y} I_{n}, \nonumber \\
    b^*_{y} &=& b_y + \half (y-\Phi \hat{w}_y)' W_{y} (y-\Phi \hat{w}_y), \nonumber \\
     \Lambda_y & = &  \Phi' \Sigma^{-1}_y \Phi, \nonumber \\
 \Lambda_\eta & = & \lambda_\eta \Phi' \Phi, \\
 \Sigma_{w_y} &=& \mbox{diag}(\lambda_{w_1}^{-1}, \ldots, \lambda_{w_q}^{-1}), \nonumber \\
 \Sigma_{w_y w} &=& \begin{pmatrix}
     \lambda^{-1}_{w1}R(\theta,\bt^*;\rho_{1}) & 0 & 0 \\
     0 & \ddots & 0 \\
     0 & 0 & \lambda^{-1}_{wq}R(\theta,\bt^*;\rho_{q})
     \end{pmatrix}, \nonumber \\
     &&\mbox{where $R(\theta,\bt^*;\rho)$ is the $1 \times m$ matrix with
                       elements $R(\theta,t^*_j;\rho)$,} \nonumber \\
     \hat{z} &=& \begin{pmatrix} \hat{w}_y \\ \hat{w} \end{pmatrix}, \nonumber \\
     \Sigma_{\hat{z}} &=& \begin{pmatrix}
       \Lambda_y^{-1} &  0 \cr
        0 & \Lambda_\eta^{-1}
     \end{pmatrix}
      +
     \begin{pmatrix}
        \Sigma_{w_y} & \Sigma_{w_y w}     \cr
        \Sigma_{w_y w}' & \Sigma_{w}
     \end{pmatrix}. \nonumber
\end{eqnarray}
The posterior distribution has the form
\begin{eqnarray}
 \label{eq:post1}
 \lefteqn{\pi(\lambda_\eta,\lambda_w,\rho,\lambda_y,
                    \theta|\hat{z}) \propto } \nonumber \\
 &&
 |\Sigma_{\hat{z}}|^{-\frac{1}{2}}
   \exp\left\{ -\half \hat{z}' \Sigma_{\hat{z}}^{-1} \hat{z} \right\}
 \times \lambda_\eta^{a^*_\eta-1} e^{-b^*_\eta \lambda_\eta}
 \times \prod_{i=1}^{q} \lambda_{wi}^{a_{w}-1} e^{-b_{w} \lambda_{wi}}
 \times \\ \nonumber
 &&
 \prod_{i=1}^{q} \prod_{k=1}^{p}
    \rho_{ik}^{a_{\rho}-1} (1- \rho_{ik})^{b_{\rho}-1}
 \times \prod_{i=1}^2 \lambda_{yi}^{a^*_y-1} e^{-b^*_y \lambda_{yi}}
 \times  I[\theta \in C],
\end{eqnarray}
where $C$ denotes the $12$-dimensional rectangle given in Table \ref{tab:params} and shown in Figure \ref{fig:design}.

\subsection{Full Bayesian GP-Based Formulation -- Combining Multiple Output
Types}
\label{sec:modcal2}

The posterior density in (\ref{eq:post1}) captures the parameter uncertainty resulting from combining the
spherical mass measurements with our statistical model formulation.  This posterior has the general
form
\[
\left\{ L(\hat{z}|\lambda_\eta,\lambda_w,\rho,\lambda_y,\theta) \cdot
\pi(\lambda_\eta) \cdot \pi(\lambda_w) \cdot \pi(\rho) \cdot \pi(\lambda_y) \right\}
\cdot \pi(\theta),
\]
where the terms within the braces are specific to this particular data type.
We can derive similar posteriors for additional data types.  These multiple data types can
be combined to inform about $\theta$ by taking the product of the terms
within the brackets
for each data type.  Hence information from $K$ data types could be combined
with the posterior
\begin{equation}
\label{eq:fullpost}
\prod_{k=1}^K \left\{ L(\hat{z}^{(k)})|\lambda_\eta^{(k)},\lambda_w^{(k)},\rho^{(k)},\lambda_y^{(k)},\theta) \cdot
\pi(\lambda_\eta^{(k)}) \cdot \pi(\lambda_w^{(k)}) \cdot \pi(\rho^{(k)}) \cdot \pi(\lambda_y^{(k)}) \right\}
\cdot \pi(\theta),
\end{equation}
where the superscript $^{(k)}$ indexes the data type.  This product form assumes independence between
error terms from different data sources.

Realizations from the posterior distribution (\ref{eq:fullpost}) are produced
by using standard, single-site MCMC. Metropolis updates \cite{Metrop:53} are
used for the components of $\rho$ and $\theta$ with a uniform proposal
distribution centered at the current value of the parameter.  The precision
parameters $\lambda_\eta$, $\lambda_w$, and $\lambda_y$ are sampled by using
Hastings updates~\cite{Hastings:70}.  Here the proposals are uniform draws,
centered at the current parameter values, with a width that is proportional to
the current parameter value.  We tune the candidate proposal width for good
Monte Carlo efficiency.
\begin{figure}[ht]
  \centerline{
   \includegraphics[width=5.5in] {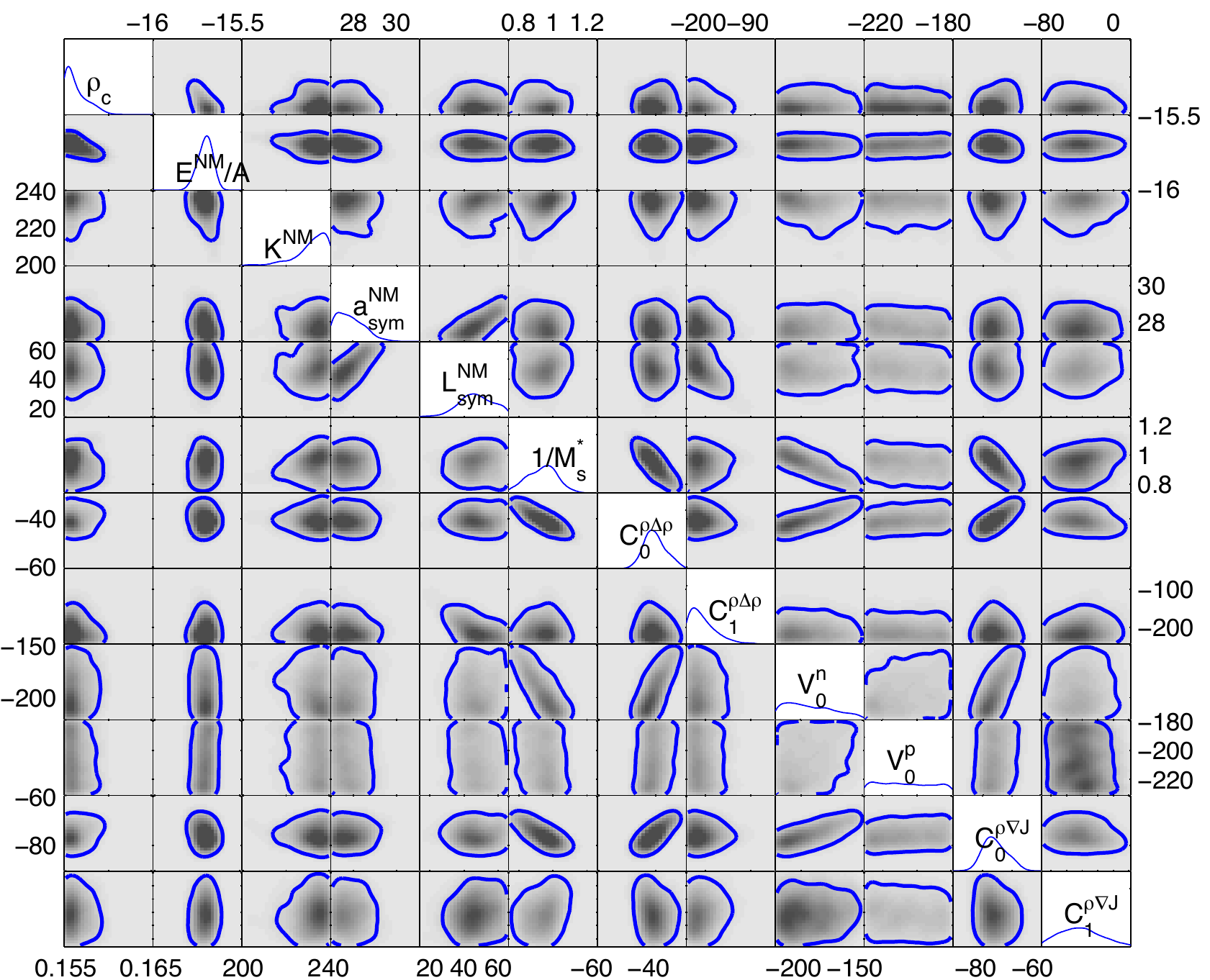}}
   \caption{\label{fig:posttheta} Univariate and bivariate marginal estimates of the
   posterior distribution for the
   12-dimensional DFT parameter vector $\theta$, conditioning on mass measurements from the spherical and deformed nuclei.
   The blue line encloses an estimated 95\% region.}
\end{figure}

\section{Posterior Results}

\begin{figure}[ht]
  \centerline  {\sf spherical nuclei \hspace{4cm} deformed nuclei \hspace{1cm}}
  \centerline{%\fbox{
   \includegraphics[height=2.5in] {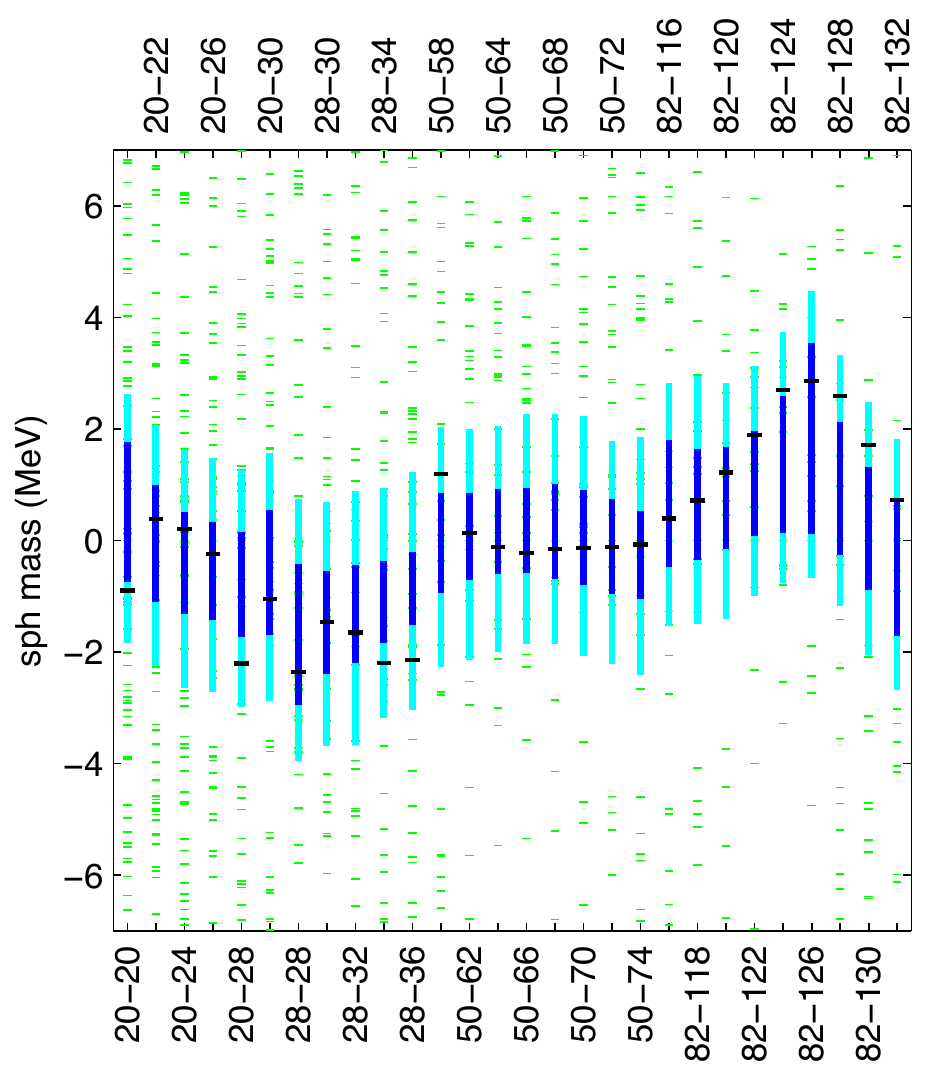}
    \includegraphics[height=2.55in] {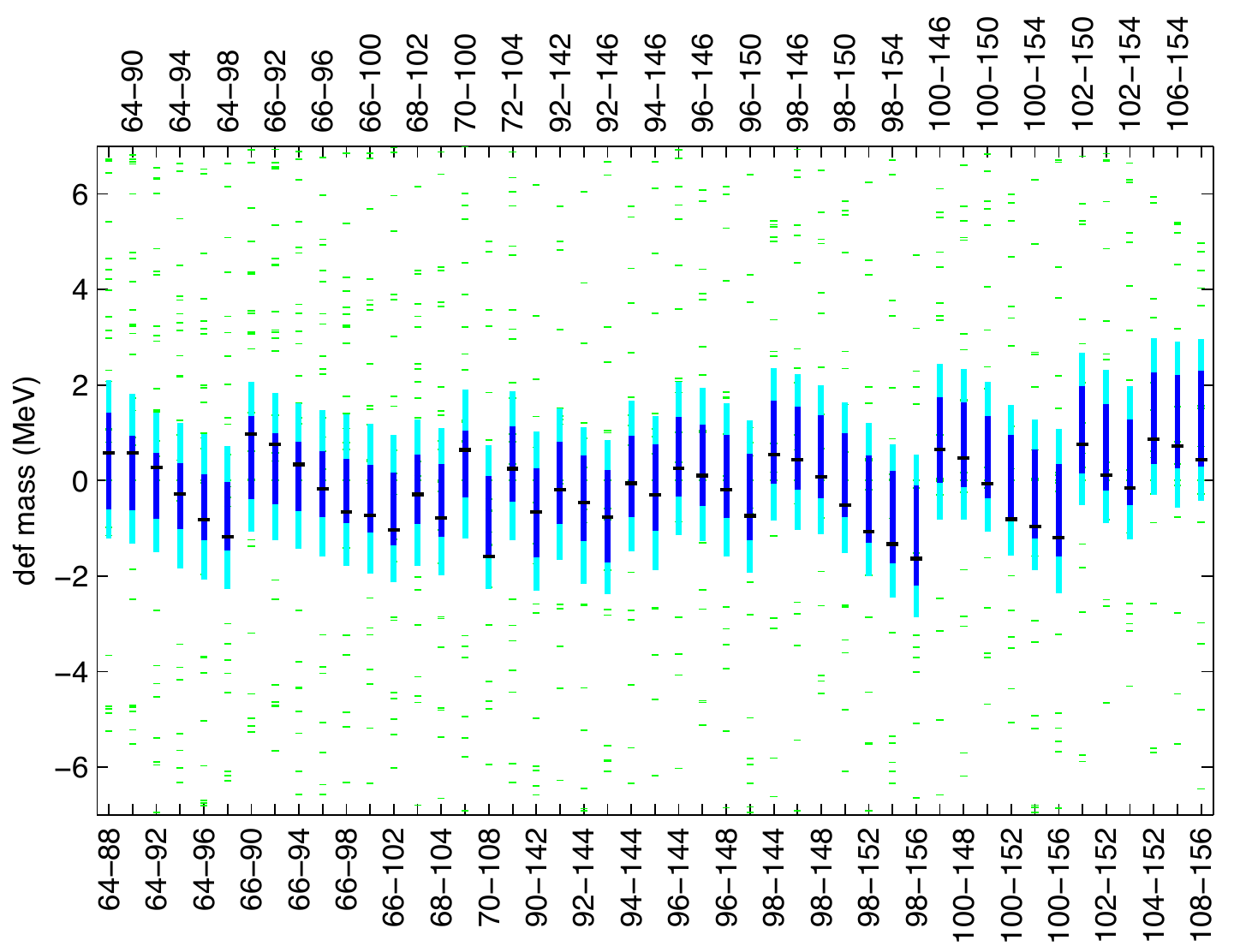}
   }%}
   \caption{\label{fig:postpred} Posterior predictions for the masses of the spherical
   and deformed nuclei used in the UNEDF1 parametrization.  The blue intervals
   correspond to $\eta(\theta)$; the light blue intervals correspond to $\eta(\theta)+\epsilon$.
   DFT model output is represented by the green dashes, and experimental
measurements by
   the black dashes. Each nucleus is labeled Z-N.}
\end{figure}

The resulting posterior distribution estimate for $\theta$ is shown in Figure  \ref{fig:posttheta} on the original scale.  The posterior values can also be propagated through the emulator to produce realizations of the model predictions.
All the predictions are centered at the nominal DFT prediction using the
parameter setting $t^0$ given
in \cite{kortelainen2012nuclear}.  Thus
the figures show the difference: $\mbox{prediction}- \eta(t^0)$.
Figure \ref{fig:postpred} shows 90\% prediction intervals for masses of the spherical and deformed nuclei used in this formulation.  The dark
blue bands show 90\% intervals for $\eta(\theta)$ -- the result of propagating uncertainty in $\theta$ through the
emulator, along with uncertainty in the
emulator.  The light blue bands show 90\% prediction intervals for the actual measured value $\eta(\theta)+\epsilon$.  This prediction
also includes the effect of uncertainty in $\epsilon$ -- the error
between $\eta(\theta)$ and $y$.  The magnitude of these errors is
controlled by $\lambda_y$, which differs for the two data types.

Comparing the posterior predictions, along with their uncertainty, with the actual mass
measurements gives an idea of how well the statistical formulation models the
experimental measurements.  The mass predictions are more accurate for the
deformed nuclei.  The dark blue parameter uncertainty bands for $\eta(\theta)$
contain all the
experimental measurements, and the estimated standard deviation of $\epsilon$ is
about 0.8 MeV.  For the spherical nuclei, 7 of the 28 measurements are not contained
in the 90\% prediction bands for $\eta(\theta)$, and the estimated standard deviation
of $\epsilon$ is about 1.3 MeV.  This larger standard deviation for $\epsilon$ is required
in order to make the experimental measurements of spherical nuclei compatible
with the statistical model formulation.
This difference in prediction quality is primarily
because the DFT model more accurately predicts the experimentally measured masses for deformed nuclei.

% \subsection{Prediction of ANL Mass Measurements}
% \label{sec:anl}

\begin{figure}[ht]
  \centerline  {\sf ANL nuclei}
  \centerline{%\fbox{
   \includegraphics[height=4.0in] {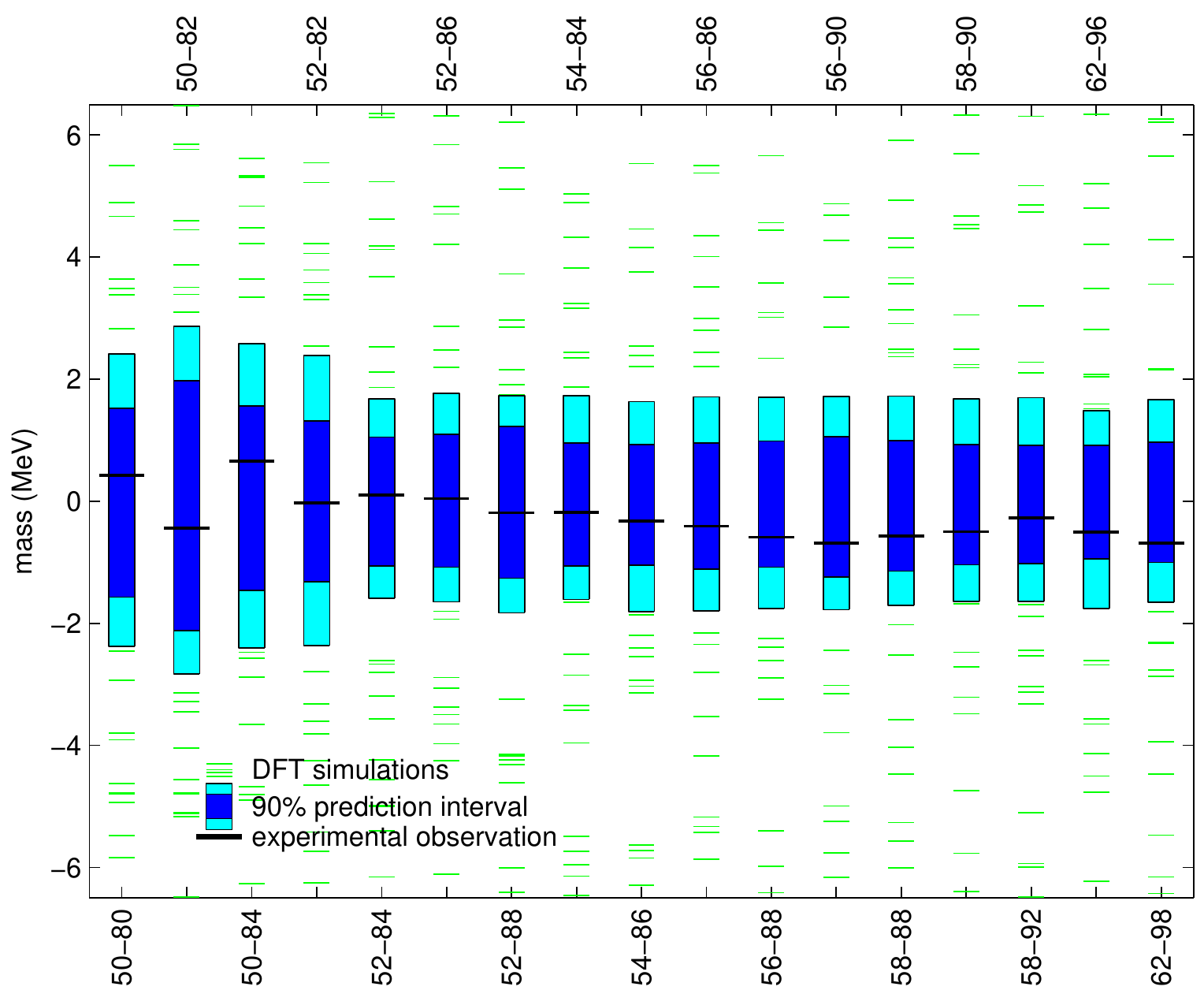}
   }%}
   \caption{\label{fig:anlpred} Posterior predictions for the 17 new ANL masses.  The blue intervals
   correspond to $\eta(\theta)$; the light blue intervals correspond to $\eta(\theta)+\epsilon$.  These
   predictions used only data from the 28 spherical and 47 deformed mass measurements used in UNEDF1.
   DFT model output is represented by the green dashes, experimental measurements by
   the black dashes. Each nucleus is labeled Z-N.}
\end{figure}

Using the posterior distribution resulting from this analysis, we can predict the outcome
of the ANL mass measurements. The posterior distribution for $\eta(\theta)^{(3)}$ for the calibrated
model can be obtained by propagating the posterior draws for $\theta$ through the ANL mass
emulator.  The resulting 90\% intervals are given by the blue intervals in Figure \ref{fig:anlpred}.
Predictions for the new measurements $\eta^{(3)}(\theta) + \epsilon^{(3)}$ require the variance, or a probabilistic
description about the variance of $\epsilon^{(3)}$ - the model error for the new ANL predictions.
For the predictions in Figure \ref{fig:anlpred} we classified each of the ANL nuclei as spherical
or deformed, and we assigned the appropriate precision estimate $\lambda_y$
obtained from
the spherical and deformed nuclei accordingly.
Since the first 4 nuclei are spherical, their prediction uncertainty is
slightly larger than that of the remaining 13 deformed nuclei.
The resulting 90\% intervals are given by the light blue lines in Figure \ref{fig:anlpred}.
%

%%%%%%%%%%%%%%%%%%%%%%%%%%%%%%%%%%%%%%%%%%%%%%%%%%%%%%%%%
%%%%%%%%%%%%%%%%%%%%%%%%%%%%%%%%%%%%%%%%%%%%%%%%%%%%%%%%%
%%%%%%%%%%%%%%%%%%%%%%%%%%%%%%%%%%%%%%%%%%%%%%%%%%%%%%%%%
%%%%%%%%%%%%%%%%%%%%%%%%%%%%%%%%%%%%%%%%%%%%%%%%%%%%%%%%%

\section{Discussion}
\label{sec:dis}

The purpose of this paper is to describe the Bayesian model calibration approach
in detail,
focusing on an example in DFT-based modeling and prediction.  While we have
treated this
statistical analysis with care, paying particular attention to
the scientific issues, this analysis
is not meant to be definitive in a scientific sense.  A more scientifically focused analysis is
given by Schunck et al.\ \cite{Schunck14} in this issue.

An important feature of this analysis is the impact of additional experimental data on the analysis.
The prediction for new measurements is given by $\eta(\theta)+\epsilon$.
In general, more data reduces the uncertainty regarding the model parameters $\theta$, reducing
the uncertainty in the calibrated model $\eta(\theta)$.  However, it will not strongly impact
the standard deviation of $\epsilon$. Hence, to produce realistic prediction
uncertainties, one must accurately characterize the uncertainty in the model
error term $\epsilon$.

How much the uncertainty in $\theta$ can be reduced by a particular type of data depends on
the DFT model $\eta(\cdot)$.  Figure \ref{fig:sens1} clearly shows that mass
data alone will not reduce
uncertainty in certain linear combinations of the model parameters.  For example, moving
$1/{M_s^*}$ and $C_0^{\rho \Delta \rho}$ together will not have a large impact on the DFT-computed
mass.  The reason is that the posterior uncertainty regarding these two
parameters is strongly correlated.
Because of this similarity in sensitivity, additional mass data will not substantially improve this situation.

This analysis estimated the error variances, allowing different precision parameters $\lambda_y$ for the
spherical and deformed masses.  This approach results in a posterior
distribution for $\theta$ that gives more
accurate results for deformed nuclei, relative to the spherical ones.  In contrast, previous analyses
\cite{kortelainen2010nuclear,kortelainen2012nuclear}
have given equal weight to the two data types.  There the weights $w$ in the objective function
effectively specify the precision $\lambda_y$ for each data type ($w = \lambda_y^{-\frac{1}{2}}$),
leading to a different posterior distribution for $\theta$.
%
%\begin{itemize}
%\item more data reduces parameter uncertainty, but increases discrepancy error.
%\item training on mass measurements gave us very useful information for the ANL mass predictions --
%  take a look at the sensitivities.
%\item this approach was applied to a wider collection of measurements (QOI's) in McDonnell et al.  In addition to
%the measurements conditioned on, we also set the discrepancy variances to match the weights used in the
%tangent model analysis of Kort et al 2012.
%\item accuracy of gp vs. tangent model.
%\item use of iid discrepancy errors may not be ideal here.  There looks to be dependence across N for a fixed Z.
%\end{itemize}

%%%%%%%%%%%%%%%%%%%%%%%%%%%%%%%%%%%%%%%%%%%%%%%%%%%%%%%%%%%%%%%%%%%%%%%%%%%%%%%
%%%%%%%%%%%%%%%%%%%%%%%%%%%%%%%%%%%%%%%%%%%%%%%%%%%%%%%%%%%%%%%%%%%%%%%%%%%%%%%

\section*{Acknowledgment}
%This material is based upon work supported by the U.S. Department of Energy, Office of Science, Office of [insert the sponsoring SC Program Office, e.g., Basic Energy Sciences]
This material was based upon work supported by the U.S.\ 
Department of Energy, Office of Science, Advanced Scientific Computing Research
SciDAC program. Computational resources were
provided through an INCITE award ``Computational Nuclear Structure'' by the
National Center for Computational Sciences (NCCS) and National Institute for
Computational Sciences (NICS) at Oak Ridge National Laboratory, through an
award by the Livermore Computing Resource Center at Lawrence Livermore National
Laboratory, and through an award by the Laboratory Computing Resource Center at
Argonne National Laboratory.

%%%%%%%%%%%%%%%%%%%%%%%%%%%%%%%%%%%%%%%%%%%%%%%%%%%%%%%%%%%%%%%%%%%%%%%%%%%%%%%
%%%%%%%%%%%%%%%%%%%%%%%%%%%%%%%%%%%%%%%%%%%%%%%%%%%%%%%%%%%%%%%%%%%%%%%%%%%%%%%
%%%%%%%%%%%%%%%%%%%%%%%%%%%%%%%%%%%%%%%%%%%%%%%%%%%%%%%%%%%%%%%%%%%%%%%%%%%%%%%
%%%%%%%%%%%%%%%%%%%%%%%%%%%%%%%%%%%%%%%%%%%%%%%%%%%%%%%%%%%%%%%%%%%%%%%%%%%%%%%

\section*{References}

\bibliographystyle{unsrt}
\bibliography{dave}

\end{document}